# THE EFFICIENCY OF SIMPLE QUANTUM ENGINE: STIRLING AND ERICSSON CYCLE.

E.O. Oladimeji[†1,2]; E.C. Umeh[1]; O.G. Abamba[3].

1. Department of Physics, Federal university Lokoja (FULOKOJA), Lokoja, Nigeria.
2. Institute of Physical Research and Technology, Peoples' Friendship University of Russia, Moscow (RUDN), Russia.
3. Department of Physics, Redeemer's University (RU), Ede, Nigeria.

## ABSTRACT

The quantum engine cycle serves as an analogous representation of the macroscopic nature of heat engines and the quantum regime of thermal devices composed of a single element. In this work, we follow the formalism of a quantum engine proposed by *Bender et al.* [1] where they observed quantum Carnot cycle with a single particle of mass $m$ confined to an infinite one-dimensional potential well of width $L$ as a working medium. Using this model, a quantum-mechanical analogue of the Stirling cycle [SC] and Ericsson cycle [EC] have been constructed through changes of both, the width of the well and its quantum state. The efficiency of quantum engines is derived, which is found to be analogous to classical thermodynamic engines.

***Keywords:*** *Quantum thermodynamics, Quantum mechanics, Ericsson cycle, Stirling cycle, Quantum heat engines, Nano-engines.*

## 1. INTRODUCTION

Classical thermodynamic engine cycles partially convert thermal energy into mechanical work, while changing pressure, temperature, and other state variables before eventually retiring to its initial state for a complete cycle [2–6]. The performance of heat engines is categorized by the efficiency and power of the cycle, this is limited to the ideal Carnot's cycle which is the maximum efficiency attainable by any reversible and cyclic engine [7–9].

Over the last four decades, divers efforts have been made to understand the relation of heat engines with quantum systems as a working mechanism which is often referred to as Quantum Heat Engines (QHEs) after being introduced in 1959 by *Scovil and Schultz-Dubois* [10]. Several working substances such as the spin systems [11–15], two-level or multilevel systems[1, 16], particle in a box [1, 17], cavity quantum electrodynamics systems[18, 19], coupled two-level systems [20], Harmonic oscillators [21–23], Pöschl-Teller Oscillator[24, 25], etc. have been observed by researchers [26, 27]. This has led to the miniaturization of classical heat cycles i.e. Joule-Brayton cycle [28–31], Otto cycle[22, 32–36], Stirling cycle [37–41], Ericsson cycles[42–44], Carnot cycle [1, 19, 45–47]etc. of which the later was used by *Bender et al* [1]where they proposed a pure-state quantum-mechanical analog of the reversible Carnot engine operating at vanishing temperatures. This engine is made of a single quantum particle confined in the one-dimensional infinite square-well potential. The role of a piston in a cylinder and temperature in classical thermodynamics was replaced with the walls of the confining potential and energy as given by the pure-state expectation value of the Hamiltonian in Quantum thermodynamics respectively. They defined the pressure $P$ (i.e, the force $F$, because of the single dimensionality) as:

$$P = -\frac{dE(L)}{dL} \tag{1}$$

---

[†] Corresponding author: E. O. Oladimeji. e-mail: nockjnr@gmail.com; enock.oladimeji@fulokja.edu.ng

In this work, we explore the application of this model in the Stirling and Ericsson cycle as shown in sections 2 and 3 respectively.

## 2. THE STIRLING CYCLE

The classical Stirling cycle is composed of two isothermal and two isochoric processes *(see Fig.1)* each of which is reversible. The effect of the regeneration at process 2 and process 4 i.e the isochoric processes are observed to have no influence in the power input. Classically, the temperature and internal energy remain constant during an isothermal process even when the system is compressed or expanded. The system remains equilibrium even when work is done. While in the quantum mechanical case, given that the system at the initial state $\psi(x)$ of volume $L$ is a linear combination of eigenstates $\phi_n(x)$ [17], the expectation value of the Hamiltonian change as the walls of the well moves. The instantaneous pressure exerted on the walls is then obtained using the relation (2). The energy value as a function of $L$ may be written as:

$$E(L) = \sum_{n=1}^{\infty} |a_n|^2 E_n \tag{2}$$

where $E_n$ is the energy spectrum (2) and the coefficients $|a_n|^2$ are limited by the normalization condition $\sum_{n=1}^{\infty} |a_n|^2 = 1$. While the isochoric process is one in which the volume $L$ of the potential well remains *constant*.

### 2.1  Process 1: Isothermal Expansion

During the Isothermal expansion, the system is excited from its initial state $n = 1$ at point 1 (i.e. from $L = L_1$ to $L = L_2$) and into the second state $n = 2$, keeping the expectation value of the Hamiltonian constant. Thus, the state of the system is a linear combination of its two energy eigenstates:

$$\Psi_n = a_1(L)\phi_1(x) + a_2(L)\phi_2(x) \tag{3}$$

where $\phi_1$ and $\phi_2$ are the wave functions of the first and second states, respectively

$$\Psi_n = a_1(L)\sqrt{2/L}\sin(\alpha x) + a_2(L)\sqrt{2/L}\sin(2\alpha x) \tag{4}$$

where $\alpha = \pi/L$. Therefore, the energy value $E(L)$ is:

$$E(L) = \sum_{n=1}^{\infty}(|a_1|^2 + |a_2|^2)E_n = |a_1|^2 E_1 + |a_2|^2 E_2 \tag{5}$$

$$E = |a_1|^2 \frac{\pi^2 \hbar^2}{2mL_1^2} + (1 - |a_1|^2)\frac{2\pi^2 \hbar^2}{mL^2} \tag{6}$$

Given that the coefficients satisfy the condition $|a_1|^2 + |a_2|^2 = 1$. The expectation value of the Hamiltonian in this state as a function of $L$ is calculated as $E = \langle \psi | H | \psi \rangle$:

$$E = \frac{\pi^2 \hbar^2}{2mL^2}(4 - 3|a_1|^2) \tag{7}$$

Setting the expectation value to be equal to $E_H$ i.e. $n = 1$

$$\frac{(4 - 3|a_1|^2)}{L^2} = \frac{1}{L_1^2} \tag{8}$$

$$L^2 = L_1{}^2(4 - 3|a_1|^2)$$

The max value of $L$ is when $L_2 = 2L_1$ and this is achieved in the isothermal expansion when $|a_1|^2 = 0$. Therefore, from eq. (8) the pressure of the isothermal expansion as a function of $L$ is:

$$P_1(L) = |a_1|^2 \frac{\pi^2 \hbar^2}{mL^3} + (1 - |a_1|^2) \frac{4\pi^2 \hbar^2}{mL^3} \tag{10}$$

Therefore;

$$P_1(L) = \frac{\pi^2 \hbar^2}{mL^3}(4 - 3|a_1|^2) \tag{11}$$

Recall that $L^2 = L_1{}^2(4 - 3|a_1|^2)$, therefore:

$$P_1(L) = \frac{\pi^2 \hbar^2}{mL_1^2 . L} \tag{12}$$

In equation (12) we can observe that the product $L.P_1(L) = constant$ and this represents that quantum analogue of the classical isothermal process [1].

### 2.2    Process 2: Isochoric Expansion

Given that the system expands isochorically from point 2 to point 3 (i.e. from $L = L_2$ to $L = L_3$) but maintain its state at $n = 2$, since no external energy is injected into the system. The expectation value of the Hamiltonian in this state as a function of $L$ is calculated as $E = \langle \psi | H | \psi \rangle$:

$$E_2 = \frac{\pi^2 \hbar^2 (2)^2}{2mL_2^2} = \frac{2\pi^2 \hbar^2}{mL_2^2} \tag{13}$$

The volume ($L$) during this process remains constant and its value is given in terms of its definition (2):

$$P = -\frac{dE}{dL} = P_2 = \frac{4\pi^2 \hbar^2}{mL_2^3} \tag{14}$$

Therefore, from eq. (4) we can express the volume $L$ in terms of the pressure $P$:

$$L_2^3 = \frac{4\pi^2 \hbar^2}{mF_2} \tag{15}$$

In eq. (15) we can observe that the product $L^3 P(L)$ is a constant and is considered the quantum analogue of the classical isochoric processes.

### 2.3    Process 3: Isothermal Compression

The system is in the second state $n = 2$ at point 3 and it compresses isothermally to the initial (ground) state $n = 1$ (i.e. from $L = L^3$ until $L = L^4$), where $L_4 = \frac{1}{2}L_3$ as the expectation value of the Hamiltonian remains constant. Thus, the pressure during the isothermal compression is:

$$P_3(L) = \frac{4\pi^2\hbar^2}{mL_3^2 \cdot L} \tag{16}$$

where the energy $E$ as a function of $L$ is $E_3 = \frac{4\pi^2\hbar^2}{2mL_3^2} = \frac{2\pi^2\hbar^2}{mL_3^2}$.

$$E_3 = \frac{2\pi^2\hbar^2}{mL_3^2} \tag{17}$$

The product $LP_3(L) = constant$. This is an exact quantum analogue of a classical *equation of state*.

### 2.4 Process 4: Isochoric Compression

The system remains at the initial state $n = 1$ at point 4 as it compresses isochorically (i.e. from $L = L_4$ until $L = L_1$). The expectation of the Hamiltonian is given by:

$$E_4 = \frac{\pi^2\hbar^2}{2mL_4^2} \tag{18}$$

and the pressure applied to the potential well's wall $P$ as a function of $L$ is:

$$P_4(L) = \frac{\pi^2\hbar^2}{mL_4^3} \tag{19}$$

$$L_4^3 = \frac{\pi^2\hbar^2}{mF_4} \tag{20}$$

The total work done $W$ by Quantum heat engine during a single closed cycle is the area of the closed-loop across the four processes as described in *(fig. 1)*. The total workdone $W$ is the sum of the work done at each process:

$$W = W_{12} + W_{23} + W_{34} + W_{41} \tag{21}$$

$$W = \int_{L_1}^{L_2} P_1(L)\, dL + \int_{L_2}^{L_3} P_2(L)\, dL + \int_{L_3}^{L_4} P_3(L)\, dL + \int_{L_4}^{L_1} P_4(L)\, dL \tag{22}$$

Therefore, the workdone is:

$$W = \int_{L_1}^{2L_1} P_1(L)\, dL + \int_{L_3}^{\frac{L_3}{2}} P_3(L)\, dL \tag{23}$$

Where the values of $P_1$ and $P_3$ are given in (12) and (16).

$$W = \int_{L_1}^{2L_1} \frac{\pi^2\hbar^2}{mL_1^2 \cdot L}\, dL + \int_{L_3}^{\frac{L_3}{2}} \frac{4\pi^2\hbar^2}{mL_3^2 \cdot L}\, dL = \frac{\pi^2\hbar^2}{mL_1^2} \ln 2 - \frac{4\pi^2\hbar^2}{mL_3^2} \ln 2 \tag{24}$$

$$W = \frac{\pi^2\hbar^2}{m}\left(\frac{1}{L_1^2} - \frac{4}{L_3^2}\right) \ln 2 \tag{25}$$

To calculate the Heat input $Q_H$ into the system, this occurred during the Isothermal expansion (process $1 \to 2$). This quantity $Q_H$ is given as:

$$Q_H = \int_{L_1}^{L_2} P_1(L) \, dL \tag{26}$$

$$Q_H = \int_{L_1}^{2L_1} \frac{\pi^2 \hbar^2}{mL_1^2 \cdot L} \, dl = \frac{\pi^2 \hbar^2}{mL_1^2} \ln 2 \tag{27}$$

By definition, the efficiency $\eta$ of a closed cycle is defined as:

$$\eta = \frac{W}{Q_H} \tag{28}$$

Where $W$ is the work done and $Q_H$ is the heat input as described in (25) and (27) respectively

$$\eta = 1 - 4\left(\frac{L_1}{L_3}\right)^2 \tag{29}$$

$$\eta = 1 - \frac{E_C}{E_H} \tag{30}$$

We can observe that this efficiency is analogous to the efficiency of a classical Stirling engine.

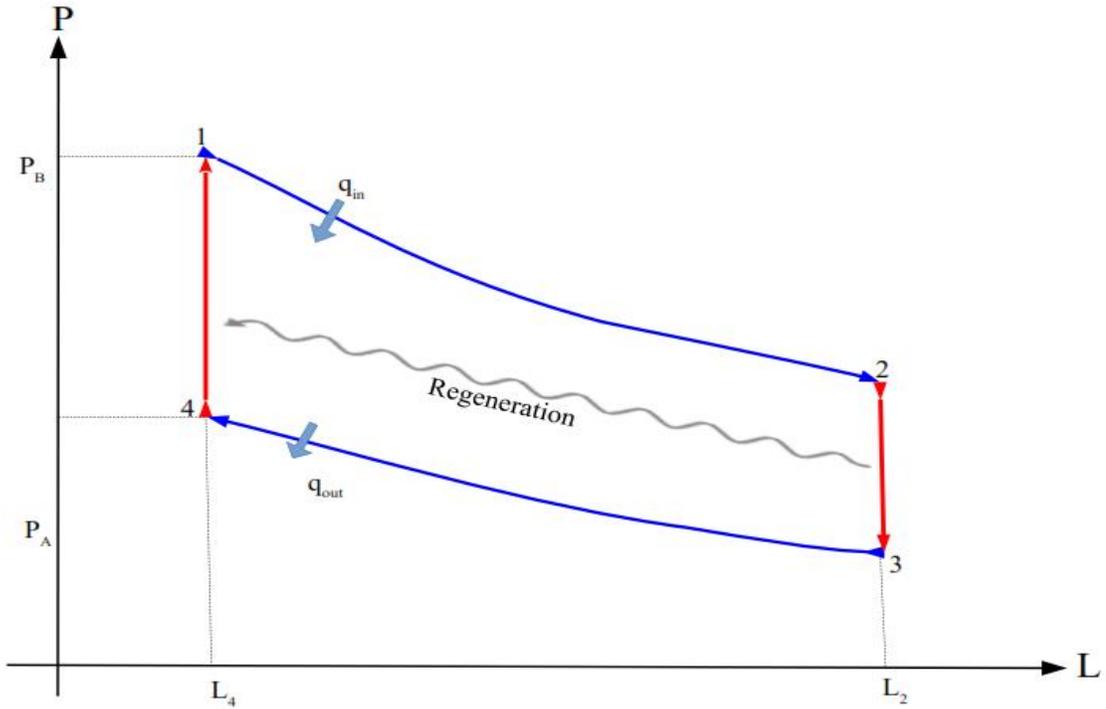

The Stirling Cycle.

Figure 1: The schematic representation of the Stirling's cycle

# 3. THE ERICSSON CYCLE

The classical Ericsson cycle undergoes two isothermal and two isobaric processes as shown in *(See Fig. 2)* each of these processes is reversible. During the isothermal process, the system's temperature remains constant while the pressure remains constant during the isobaric process.

### 3.1  Process 1: Isothermal Expansion

The classical Ericsson cycle is composed of two isothermal and two isobaric processes a cycle process as shown in *(See Fig. 2)* each of which is reversible. The isothermal process is one in which the temperature in the system is constant. During this process, heat is injected into the system which is excited from the initial state $n = 1$ at point 1 to the second state $n = 2$ at point 2. The state of the system during this process is a linear combination of the lowest two energy eigenstates. Just as derived earlier in Eq (16), the pressure of the isothermal expansion as a function of $L$ is:

$$F_1(L) = \frac{\pi^2 \hbar^2}{mL^3}(4 - 3|a_1|^2) \tag{31}$$

$L^2 = L_1^2(4 - 3|a_1|^2)$, following the same calculations in eq (8) and (9). The max $L$ is when at point $L_2$ where $|a_1|^2 = 0$.

$$F_1(L) = \frac{\pi^2 \hbar^2}{mL_1^2 \cdot L} \tag{32}$$

The product $L \cdot P_1(L) = constant$, this represents that quantum analogue of the classical isothermal process [1]

### 3.2  Process 2: Isobaric Expansion

During the process $2 \rightarrow 3$, the system expands isobarically from $L = L_2$ until $L = L_3$ the heat input is stopped as no energy enters the system so that the particle remains in the second state $n = 2$ and the change in the internal energy is equivalent to the work done against the walls of the well. The expectation value of the Hamiltonian is $E = \frac{\pi^2 \hbar^2}{2mL^2}$, and the pressure is given by

$$P(L) = \frac{\pi^2 \hbar^2}{mL^3} \tag{33}$$

or $\frac{E}{L} = \frac{\pi^2 \hbar^2}{2mL_2^3} = constant$, which represents the isobaric analogous equation.

### 3.3  Process 3: Isothermal Compression

During the process $3 \rightarrow 4$, the system compresses isothermally from $L = L_3$ until $L = L_4$. During the process, $L_4 = 2L_3$ is the expectation value of the Hamiltonian remains constant as the system returns to its ground state $n = 1$ from its excited state $n = 2$.

$$E_1 = \frac{\pi^2 \hbar^2}{2mL_3^2} = E_H \tag{34}$$

$$P_1(L) = \frac{\pi^2 \hbar^2}{mL_3^2 \cdot L} \tag{35}$$

### 3.4 Process 4: Isobaric Compression

During the process $4 \to 1$, the system compresses isobarically from $L = L_4$ until $L = L_1$ the heat input is stopped as no energy enters the system so that the particle remains at the ground state $n = 1$. The expectation value of the Hamiltonian is $E = \frac{4\pi^2 \hbar^2}{2mL^2} = \frac{2\pi^2 \hbar^2}{mL_3^2}$, and the pressure is given by

$$P(L) = \frac{4\pi^2 \hbar^2}{mL^3} \tag{36}$$

As shown earlier in eq (22) the work performed by the system during one cycle along the four processes is defined as:

$$W = \int_{L_1}^{L_2} F_2(L)\, dL + \int_{L_2}^{L_3} F(L)\, dL + \int_{L_3}^{L_4} F_1(L)\, dL + \int_{L_4}^{L_1} F(L)\, dL$$

Recall that, $L_2 = \frac{1}{2}L_1$ and $L_4 = 2L_3$. Therefore;

$$W = \int_{L_1}^{\frac{L_1}{2}} F_2(L)\, dL + \int_{\frac{L_1}{2}}^{L_3} F(L)\, dL + \int_{L_3}^{2L_3} F_1(L)\, dL + \int_{2L_3}^{L_1} F(L)\, dL \tag{37}$$

$$W = \int_{L_1}^{\frac{L_1}{2}} \frac{4\pi^2 \hbar^2}{mL_1^2 \cdot L}\, dL + \int_{\frac{L_1}{2}}^{L_3} \frac{\pi^2 \hbar^2}{mL^3}\, dL + \int_{L_3}^{2L_3} \frac{\pi^2 \hbar^2}{mL_3^2 \cdot L}\, d + \int_{2L_3}^{L_1} \frac{4\pi^2 \hbar^2}{mL^3}\, dL \tag{38}$$

$$W = -\frac{4\pi^2 \hbar^2}{mL_1^2} \ln 2 + \frac{\pi^2 \hbar^2}{mL_3^2} \ln 2$$

$$W = \frac{\pi^2 \hbar^2}{m} \left( \frac{1}{L_3^2} - \frac{4}{L_1^2} \right) \ln 2 \tag{39}$$

The efficiency $\eta$ of the heat engine is defined to be:

$$\eta = \frac{W}{Q_H} \tag{40}$$

given that $Q_H$ is the quantity of heat in the high-temperature reservoir and $W$ is the work done by the classical heat engine. Where $Q_H$ is the heat engine injected into the potential well during the isothermal expansion in a quantum engine:

$$Q_H = \frac{\pi^2 \hbar^2}{mL_3^2} \ln 2 \tag{41}$$

Therefore, the efficiency $\eta$ of the quantum heat engine is:

$$\eta = 1 - 4 \left( \frac{L_3}{L_1} \right)^2 \tag{42}$$

Substituting the eqs. (40) and (41) into (42), the efficiency can be written as:

$$\eta = 1 - \frac{E_C}{E_H} \tag{43}$$

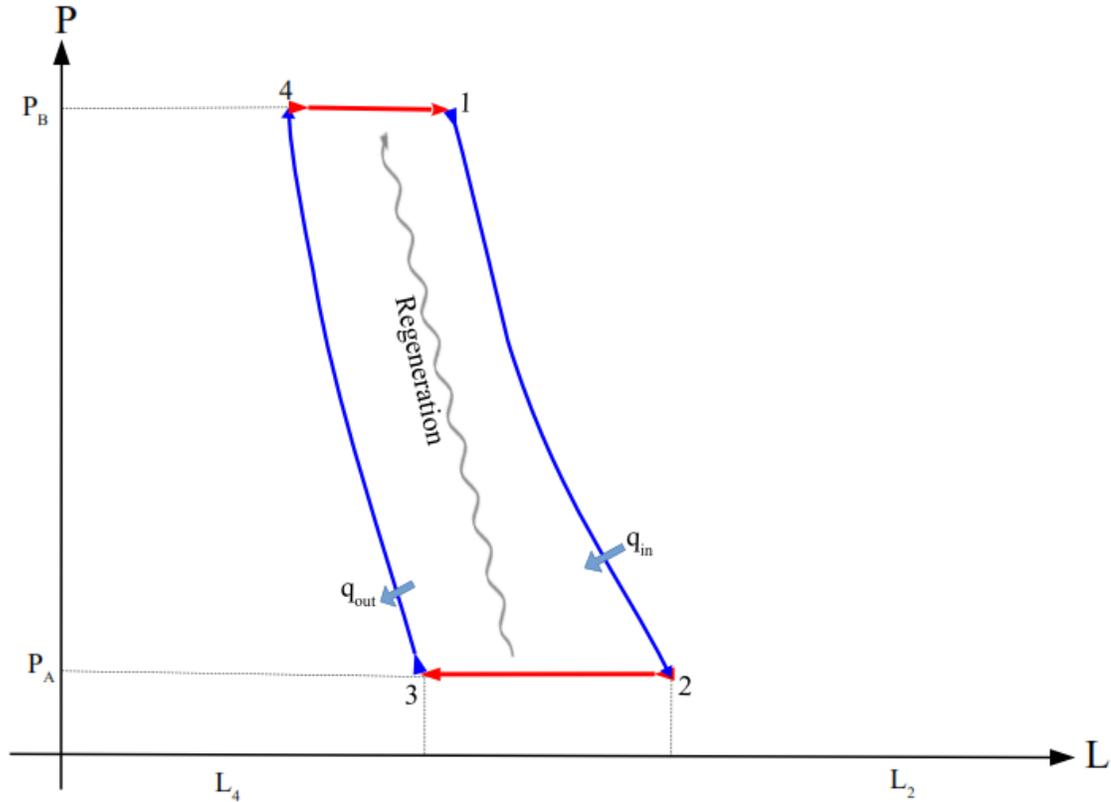

The Ericsson Cycle.

*Figure 2: The schematic representation of the Ericsson's cycle*

Our derived efficiencies in eq(30) and (35) respectively are analogous to the classical Stirling and Ericsson cycles.

## OUR RESULT

As stated earlier, the performance of an engine is determined by its efficiency and power. In this work, we applied the quantum mechanical particles confined in a potential well as working fluid to Stirling and Ericsson cycles. We observed the quantum interpretation the isothermal and isochoric process in the stirling cycle likewise the isothermal and isobaric process in the ericsson cycle. We showed that the efficiencies of these two quantum cycles are analogous to their counterparts from classical thermodynamics.

# REFERENCE

1. Bender, C.M., Brody, D.C., Meister, B.K.: Quantum mechanical Carnot engine. J. Phys. A. Math. Gen. 33, 4427–4436 (2000). https://doi.org/10.1088/0305-4470/33/24/302

2. Callen, H.B.: Thermodynamics and an Introduction to Thermostatics. John Wiley & Sons, New York (1985)

3. Yunus A, C., Michael A, B.: Thermodynamics An Engineering Approach. McGraw-Hill Education (2015)

4. Astarita, G.: Thermodynamics An Advanced Textbook for Chemical Engineers. , New-York (1990)

5. Abah, O., Lutz, E.: Energy efficient quantum machines. EPL. 118, 40005 (2017). https://doi.org/10.1209/0295-5075/118/40005

6. Abah, O., Roßnagel, J., Jacob, G., Deffner, S., Schmidt-Kaler, F., Singer, K., Lutz, E.: Single-Ion Heat Engine at Maximum Power. Phys. Rev. Lett. 109, 203006 (2012). https://doi.org/10.1103/PhysRevLett.109.203006

7. Stark, J.P.: Fundamentals of classical thermodynamics (Van Wylen, Gordon J.; Sonntag, Richard E.). J. Chem. Educ. 43, A472 (1966). https://doi.org/10.1021/ed043pA472.1

8. Wood, B.D.: Applications of Thermodynamics. Addison-Wesley Publishing Company (1982)

9. Spalding, D.B., Cole, E.H.: Engineering Thermodynamics: D.B. Spalding, E.H. Cole. Edward Arnold (1966)

10. Scovil, H.E.D., Schulz-DuBois, E.O.: Three-Level Masers as Heat Engines. Phys. Rev. Lett. 2, 262–263 (1959). https://doi.org/10.1103/PhysRevLett.2.262

11. Feldmann, T., Kosloff, R.: Performance of discrete heat engines and heat pumps in finite time. Phys. Rev. E. 61, 4774–4790 (2000). https://doi.org/10.1103/physreve.61.4774

12. Feldmann, T., Kosloff, R.: Quantum four-stroke heat engine: Thermodynamic observables in a model with intrinsic friction. Phys. Rev. E - Stat. Physics, Plasmas, Fluids, Relat. Interdiscip. Top. 68, 18 (2003). https://doi.org/10.1103/physreve.68.016101

13. Feldmann, T., Kosloff, R.: Characteristics of the limit cycle of a reciprocating quantum heat engine. Phys. Rev. E - Stat. Physics, Plasmas, Fluids, Relat. Interdiscip. Top. 70, 13 (2004). https://doi.org/10.1103/physreve.70.046110

14. Wu, F., Chen, L., Wu, S., Sun, F., Wu, C.: Performance of an irreversible quantum Carnot engine with spin 1/2. J. Chem. Phys. 124, 214702 (2006). https://doi.org/10.1063/1.2200693

15. Wang, J., He, J., Xin, Y.: Performance analysis of a spin quantum heat engine cycle with internal friction. Phys. Scr. (2007). https://doi.org/10.1088/0031-8949/75/2/018

16. Quan, H.T., Liu, Y.X., Sun, C.P., Nori, F.: Quantum thermodynamic cycles and quantum heat engines. Phys. Rev. E. 76, 31105 (2007). https://doi.org/10.1103/PhysRevE.76.031105

17. Guzmán-Vargas, L., Granados, V., Mota, R.D.: Efficiency of simple quantum engines: The Joule-Brayton and Otto cycles. AIP Conf. Proc. 643, 291–296 (2002). https://doi.org/10.1063/1.1523819

18. Scully, M.O., Zubairy, M.S., Agarwal, G.S., Walther, H.: Extracting Work from a Single Heat Bath via Vanishing Quantum Coherence. Science (80-. ). 299, 862–864 (2003).

https://doi.org/10.1126/science.1078955

19. Quan, H.T., Zhang, P., Sun, C.P.: Quantum-classical transition of photon-Carnot engine induced by quantum decoherence. Phys. Rev. E. 73, 036122 (2006). https://doi.org/10.1103/PhysRevE.73.036122

20. Henrich, M.J., Mahler, G., Michel, M.: Driven spin systems as quantum thermodynamic machines: Fundamental limits. Phys. Rev. E. 75, 051118 (2007). https://doi.org/10.1103/PhysRevE.75.051118

21. Tah, R.: Simulating a Quantum Harmonic Oscillator by introducing it to a Bosonic System. (2020)

22. Kosloff, R., Rezek, Y.: The Quantum Harmonic Otto Cycle. Entropy. 19, 1–36 (2017). https://doi.org/10.3390/e19040136

23. Geva, E., Kosloff, R.: A quantum-mechanical heat engine operating in finite time. a model consisting of spin-1/2 systems as the working fluid. J. Chem. Phys. 96, 3054–3067 (1992). https://doi.org/10.1063/1.461951

24. Oladimeji, E.O. The efficiency of quantum engines using the Pöschl – Teller like oscillator model. Phys. E Low-Dimensional Syst. Nanostructures. 111, 113–117 (2019). https://doi.org/10.1016/j.physe.2019.03.002

25. Oladimeji, E.O., Owolabi, S.O., Adeleke, J.T.: The Pöschl-Teller Like Description of Quantum-Mechanical Carnot Engine. (2019)

26. Wang, J., He, J., He, X.: Performance analysis of a two-state quantum heat engine working with a single-mode radiation field in a cavity. Phys. Rev. E - Stat. Nonlinear, Soft Matter Phys. 84, 1–6 (2011). https://doi.org/10.1103/PhysRevE.84.041127

27. Peng, H.-P., Fang, M.-F., Zhang, C.-Y.: Quantum Heat Engine Based on Working Substance of Two Particles Heisenberg XXX Model with the Dzyaloshinskii-Moriya Interaction. Int. J. Theor. Phys. 58, 1651–1658 (2019). https://doi.org/10.1007/s10773-019-04061-3

28. Açikkalp, E., Caner, N.: Application of exergetic sustainability index to a nano-scale irreversible Brayton cycle operating with ideal Bose and Fermi gasses. Phys. Lett. Sect. A Gen. At. Solid State Phys. 379, 1990–1997 (2015). https://doi.org/10.1016/j.physleta.2015.06.020

29. Toro, C., Lior, N.: Analysis and comparison of solar-heat driven Stirling, Brayton and Rankine cycles for space power generation. Energy. (2017). https://doi.org/10.1016/j.energy.2016.11.104

30. Sahin, B., Kodal, A., Yavuz, H.: Efficiency of a joule-brayton engine at maximum power density. J. Phys. D. Appl. Phys. (1995). https://doi.org/10.1088/0022-3727/28/7/005

31. Liu, X., Chen, L., Ge, Y., Wu, F., Sun, F.: Fundamental optimal relation of a spin 1/2 quantum Brayton heat engine with multi-irreversibilities. Sci. Iran. 19, 1124–1132 (2012). https://doi.org/10.1016/j.scient.2012.06.013

32. Chen, J.F., Sun, C.P., Dong, H.: Boosting the performance of quantum Otto heat engines. Phys. Rev. E. 100, (2019). https://doi.org/10.1103/PhysRevE.100.032144

33. He, J., He, X., Tang, W.: The performance characteristics of an irreversible quantum Otto harmonic refrigeration cycle. Sci. China, Ser. G Physics, Mech. Astron. (2009). https://doi.org/10.1007/s11433-009-0169-z

34. Türkpençe, D., Altintas, F.: Coupled quantum Otto heat engine and refrigerator with inner friction. Quantum Inf. Process. 18, (2019). https://doi.org/10.1007/s11128-019-2366-7

35. Çakmak, S., Türkpençe, D., Altintas, F.: Special coupled quantum Otto and Carnot cycles. Eur. Phys. J. Plus. 132, (2017). https://doi.org/10.1140/epjp/i2017-11811-3

36. Latifah, E., Purwanto, A.: Multiple-state quantum Otto engine, 1D box system. In: AIP Conference Proceedings. pp. 137–140. American Institute of Physics Inc. (2014)

37. Huang, X.L., Niu, X.Y., Xiu, X.M., Yi, X.X.: Quantum Stirling heat engine and refrigerator with single and coupled spin systems. Eur. Phys. J. D. 68, (2014). https://doi.org/10.1140/epjd/e2013-40536-0

38. Izrailovich, M.Y., Sinev, A. V., Shcherbakov, V.F., Kangun, R. V.: A dynamic model of a self-vibration cycle in a stirling engine with opposed cylinders. J. Mach. Manuf. Reliab. 36, (2007). https://doi.org/10.3103/s1052618807030028

39. Yin, Y., Chen, L., Wu, F.: Performance of quantum Stirling heat engine with numerous copies of extreme relativistic particles confined in 1D potential well. Phys. A Stat. Mech. its Appl. 503, (2018). https://doi.org/10.1016/j.physa.2018.02.202

40. Yin, Y., Chen, L., Wu, F.: Optimal power and efficiency of quantum Stirling heat engines. Eur. Phys. J. Plus. 132, (2017). https://doi.org/10.1140/epjp/i2017-11325-0

41. Torres García, M., Carvajal Trujillo, E., Vélez Godiño, J., Sánchez Martínez, D.: Thermodynamic Model for Performance Analysis of a Stirling Engine Prototype. Energies. 11, 2655 (2018). https://doi.org/10.3390/en11102655

42. Sisman, A., Saygin, H.: On the power cycles working with ideal quantum gases: I. The Ericsson cycle. J. Phys. D. Appl. Phys. 32, 664–670 (1999). https://doi.org/10.1088/0022-3727/32/6/011

43. Fula, A., Stouffs, P., Sierra, F.: In-cylinder heat transfer in an ericsson engine prototype. Renew. Energy Power Qual. J. 1, (2013). https://doi.org/10.24084/repqj11.594

44. Bonnet, S., Alaphilippe, M., Stouffs, P.: Energy, exergy and cost analysis of a micro-cogeneration system based on an Ericsson engine. Int. J. Therm. Sci. 44, (2005). https://doi.org/10.1016/j.ijthermalsci.2005.09.005

45. Liu, X., Chen, L., Wu, F., Sun, F.: Ecological optimization of an irreversible quantum Carnot heat engine with spin-1/2 systems. Phys. Scr. (2010). https://doi.org/10.1088/0031-8949/81/02/025003

46. Jaynes, E.T.: The Evolution of Carnot's Principle. In: Maximum-Entropy Bayesian Methods Sci. Eng. pp. 267–281 (1988)

47. Abe, S.: Maximum-power quantum-mechanical Carnot engine. Phys. Rev. E - Stat. Nonlinear, Soft Matter Phys. 83, 1–3 (2011). https://doi.org/10.1103/PhysRevE.83.041117